\documentclass[%
nofootinbib,reprint,
amsmath,amssymb,
aps,
]{article}
\usepackage{easyReview}
\usepackage{graphicx}
\usepackage{dcolumn}
\usepackage{bm}
\usepackage{color}
\usepackage{amsmath}
\usepackage{epsfig,graphics}
\usepackage{etoolbox}
\usepackage{gensymb}
\usepackage{slashed}
\usepackage{jheppub}
\usepackage[capitalise]{cleveref}
\usepackage{multirow}
\usepackage{physics}
\usepackage{cleveref}
\usepackage{subfigure}
\usepackage{comment}
\definecolor{red}{rgb}{0.9, 0,0}
\definecolor{defgrey}{HTML}{9f9f9f}
\definecolor{cerulean}{rgb}{0., 0.62,0.9}
\definecolor{navy}{rgb}{0.05, 0.05,0.8}
\definecolor{orange}{rgb}{1., 0.65,0.0}
\definecolor{magenta}{rgb}{1.0, 0.08, 0.58}
\definecolor{green}{rgb}{0.1, 0.5, 0.19}
\definecolor{amber}{rgb}{1.0, 0.75, 0.0}
\definecolor{blond}{rgb}{0.98, 0.94, 0.75}
\definecolor{median}{HTML}{FFCC99}
\definecolor{mean}{HTML}{EEDC82}

\newcommand{\GeV}{\text{ GeV}}

\newcommand{\BelleII}{Belle~II }

\newcommand{\gagg}{g_{a\gamma\gamma}}

\newcommand{\madgraph}{\textsc{Madgraph5\_aMC@NLO}}
\def\eq#1{Eq.~(\ref{#1})}
\newcommand{\tplus}{t_{+}}
\newcommand{\tminus}{t_{-}}
\newcommand{\sminus}{s_{-}}
\newcommand{\splus}{s_{+}}

\makeatletter
\patchcmd\@footnotemark{{link}}{{footnote}}{}{\fail}
\makeatother
\ActivateGenericHook{hyp/link/footnote}
\AddToHook{hyp/link/footnote}{\hypersetup{linkcolor=cyan}}
\begin{document}

\title{Fusing photons into diphoton resonances at Belle II and beyond}

\author{Francesca Acanfora$^{*\dagger}$} 
\author{Roberto Franceschini$^\dagger$} 
\author{Alessio Mastroddi$^\dagger$}
\author{Diego Redigolo$^{\ddagger}$}
\affiliation{
$^*$ Institute for Theoretical Particle Physics (TTP), Karlsruhe Institute of Technology (KIT), D-76131 Karlsruhe, Germany \\
$^\dagger $ Universit\`a degli Studi and INFN Roma Tre, Via della Vasca Navale 84, I-00146, Rome, Italy\\
$^\ddagger$ INFN, Sezione di Firenze Via G. Sansone 1, 50019 Sesto Fiorentino, Italy, \\
}

\abstract{
We propose a new search for a diphoton resonance in the $e^+e^-+\gamma\gamma$ final state at Belle~II that improves the expected reach compared to the $\gamma+\gamma\gamma$ channel in  most of the available mass range. For simplicity we show our results in the simple parameter space of an ALP coupled solely to Standard Model photons. 
In addition, we show how an extension of the forward coverage of Belle~II, or another similar experiment at the high intensity frontier, could improve the reach in our channel. We show that such a forward extension can be advantageous even with a loss of a factor 100 in luminosity  compared to  Belle~II.}    

\maketitle

\section{Introduction}\label{sec:introduction}

Diphoton resonances have been an extremely important final state in the history of particle physics, which culminated with the Standard Model Higgs boson discovery at ATLAS and CMS~\cite{CMS:2012qbp,ATLAS:2012yve}. Without any doubt, finding a new peak in the diphoton invariant mass distribution will give a decisive hint of new physics. For this reason, independently on the theory motivations, it is of utmost importance to think about new ways to enhance the sensitivity of diphoton resonance searches at existing colliders. This program has been pursued in the recent years in different directions and lead to a significant extension of the reach on diphoton resonances for different production channels, with a special emphasis in the relatively unexplored region of low invariant mass for the diphoton pair~\cite{Knapen:2016moh,Knapen:2017ebd,Mariotti:2017vtv,CidVidal:2018blh,Aloni:2019ruo,Wang:2021uyb,Knapen:2021elo,ATLAS:2022abz,Alonso-Alvarez:2023wni,CMS:2018erd,Balkin:2023gya}. 

\begin{figure*}[t!]
\centering
\includegraphics[width=0.69 \columnwidth]{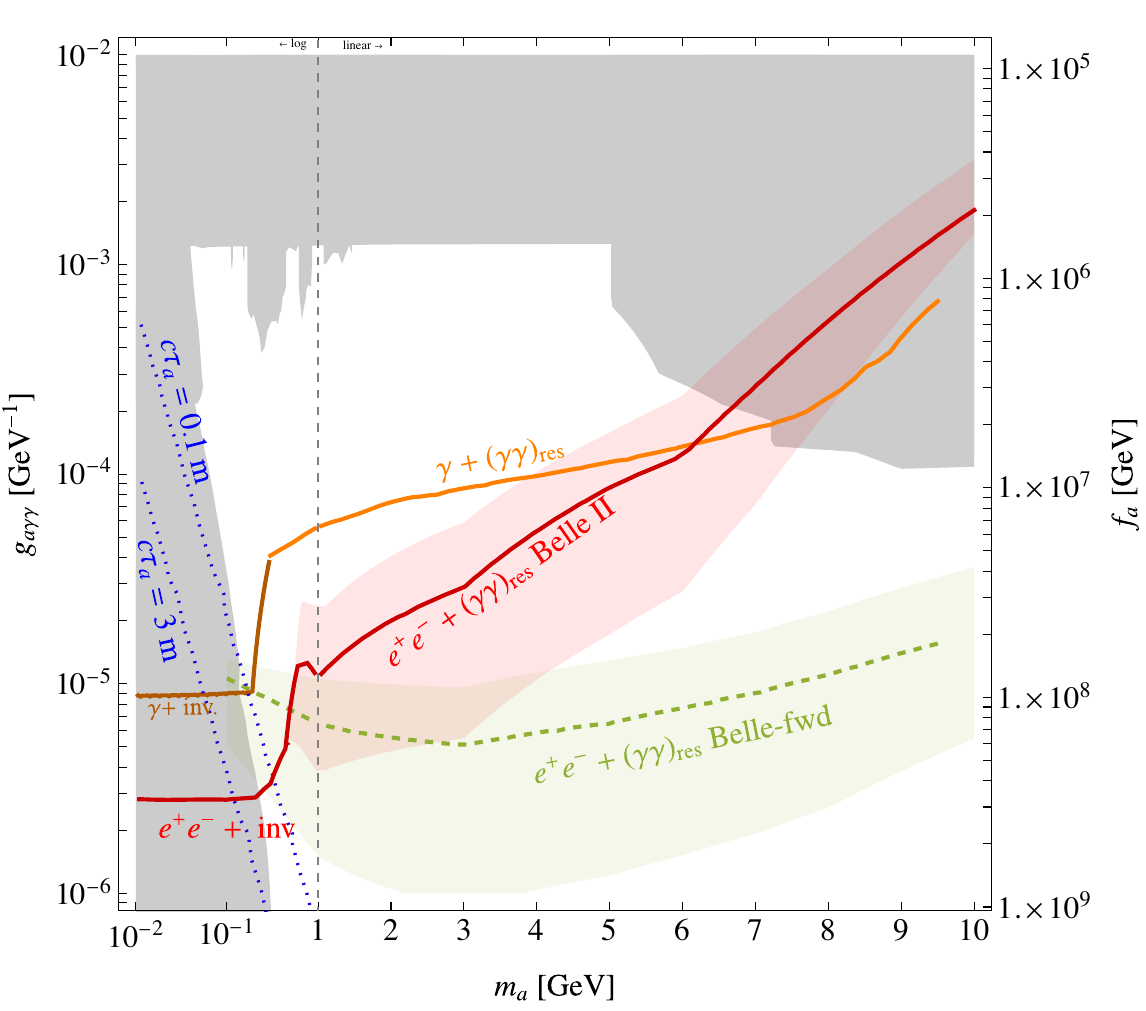}
\caption{Expected sensitivity of \BelleII at 95\% C.L. to the ALP coupling to photons $g_{a\gamma\gamma}$ as defined in ~\eq{eq:ALPlagrangian}. In {\color{red}\bf red} we show the expected reach of $e^+e^-+(\gamma\gamma)_{\text{res}}$  derived here and in {\color{orange}\bf orange} the one of $\gamma+(\gamma\gamma)_{\text{res}}$ derived in Ref.~\cite{Dolan:2017osp} both assuming the present Belle II detector coverage (see Eq.~\eqref{eq:BelleIIopen}) and $50\text{ ab}^{-1}$ of integrated luminosity. The details of our analysis are given in Sec.~\ref{sec:Belle2}. The {\color{green}\bf green dashed} line corresponds to the reach of a hypothetical experiment with increased forward acceptance (see Eq.~\eqref{eq:bellefwd}) and identical luminosity. The details are given in  Sec.~\ref{sec:forwardfactory}. The {\color{red}\bf red error band} and the {\color{green}\bf green error band} on both expected reaches correspond to the different statistical treatment of the empty bins in  the four-dimensional binned likelihood, as detailed in App.~\ref{sec:MC} and App.~\ref{sec:likelihood}. The {\color{navy}\bf dotted blue} lines are  isolines for the sample averaged ALP decay length measured in the lab-frame. When the ALP decays become long enough, the expected sensitivity of $e^+e^-+\text{invisible}$~\cite{Acanfora:2023gzr} and of $\gamma+\text{invisible}$~\cite{Dolan:2017osp} are also shown. The {\color{defgrey}\bf gray shaded} region shows existing constraints, assuming $g_{a\gamma\gamma}$ is generated above the electroweak scale by an ALP coupling to the $U(1)_Y$ gauge boson (see Ref.~\cite{Dolan:2017osp,Ferber:2022rsf} for the case where the ALP couples to $SU(2)$ gauge bosons). These include limits from beam dump experiments~\cite{Riordan:1987aw,Dobrich:2015jyk}, LEP and LHC limits recast in Ref.~\cite{Knapen:2016moh}, constraints from PrimEX data derived in Ref.~\cite{Aloni:2019ruo}, the recent limits from the Belle~II search on $\gamma+(\gamma\gamma)_{\text{res}}$~\cite{Belle-II:2020jti}, from BES~III data~\cite{BESIII:2022rzz} and from ATLAS~\cite{Knapen:2017ebd} and CMS~\cite{CMS:2018erd} searches on ultra-peripheral heavy ions collisions.}\label{fig:money} 
\end{figure*}

In this paper, we propose a new search for a diphoton resonance at Belle~II taking advantage of the ``photon fusion'' channel
\begin{equation}
e^{+}e^{- }\to e^+e^-+(\gamma\gamma)_{\text{res}}\,, \label{eq:signal-intro}
\end{equation}
where $(\gamma\gamma)_{\text{res}}$ indicates the diphoton resonant pair. In concrete scenarios where a new scalar (or pseudoscalar) is coupled to the SM photons, we show how this channel can improve the reach with respect to the more commonly considered ``ALP-strahlung'' channel~\cite{Dolan:2017osp} 
\begin{equation}
e^{+}e^{-} \to \gamma+(\gamma\gamma)_{\text{res}} \,.
\end{equation}

Despite the smaller production cross-section with respect to $e^+e^-\to\gamma+(\gamma\gamma)_{\text{res}}$, the  $e^+e^-\to e^+e^-+(\gamma\gamma)_{\text{res}}$ process has characteristic features that make this signal highly distinguishable from the QED background. Ultimately, these features stem from the higher dimensional observable phase-space with respect to the ``ALP-strahlung'' process. 

For instance, in the SM process the photons are preferably emitted at small angle with respect to the $e^\pm$ which radiated them and tend to carry little energy, while the signal allows for a large angular separation and energetic photons. Moreover, at low invariant masses the photons in the SM process tend to carry very little energy compared to the signal ones. In the following, we will elaborate further on the discrimination that can be attained depending on the invariant mass of the diphoton resonance.



For concreteness, we consider the simple model of an axion-like particle (ALP) coupled to SM photons,  but we expect that our findings can be applied to a much broader class of diphoton resonances. The ALP Lagrangian that we consider is  
\begin{equation}
\mathcal{L}=\frac{1}{2}(\partial_\mu a)^2 -\frac{m_a^2}{2} a^2 - \frac{g_{a\gamma\gamma}}{4} a F_{\mu\nu}\tilde{F}^{\mu\nu}\ \label{eq:ALPlagrangian}\ ,
\end{equation}
and in Fig.~\ref{fig:money} we summarize the expected reach on $g_{a\gamma\gamma}$ at Belle~II with $50\text{  ab}^{-1}$ of luminosity. The red line shows the projected sensitivity of our proposed  $e^+e^-+(\gamma\gamma)_{\text{res}}$ channel in the ALP parameter space compared with the one of the $e^+e^-\to\gamma+(\gamma\gamma)_{\text{res}}$ channel derived in Ref.~\cite{Dolan:2017osp}. The improvement of the reach in $g_{a\gamma\gamma}$ is substantial for an ALP lighter than 6 GeV promptly decaying into di-photons. For higher masses, the ALP searches at Belle~II quickly become less sensitive than existing constraint from ultraperipheral heavy ion collisions at ATLAS~\cite{Knapen:2017ebd} and CMS~\cite{CMS:2018erd}.
In Sec.~\ref{sec:Belle2} we present the details of the analysis that lead to the sensitivity curve at Belle~II of Fig.~\ref{fig:money}.

Given the ALP life-time  
\begin{equation}
c\tau_a\simeq  1.5\text{ mm}\left[\frac{1\text{ GeV}}{m_a}\right]^3\left[\frac{5\times10^{-6} \text{ GeV}^{-1}}{g_{a\gamma\gamma}}\right]^2\,, 
\end{equation}
for sufficiently light ALP masses the decay length becomes macroscopic, ultimately exceeding the length of the \BelleII decay volume. For this reason, the expected sensitivity in Fig.~\ref{fig:money} accounts also for the possibility that the ALP decay happens so far out from the interaction point its production leaves no trace in the detectors, resulting in a $e^+e^-+\text{invisible}$ final state. We studied this signal in our earlier Ref.~\cite{Acanfora:2023gzr}. The details of how this effect is taken into account in our simulation are given in App.~\ref{app:lifetime}. In the intermediate region, a dedicated search for displaced diphoton resonance could possibly further enhance the sensitivity. Here we do not distinguish prompt or displaced diphoton resonance and ignore any possible quality cut on the reconstruction of the diphoton vertex so that any displaced diphoton event is considered as prompt.\footnote{In absence of such dedicated analysis, our sensitivity estimate suffers a small, but   appreciable, drop in the mass range $m_a\simeq 0.1-1\text{ GeV}$. For  masses around these values and for the coupling strength at which we can put a bound the decay length becomes comparable to the $\mathcal{O}(m)$ size of the detector, resulting in a loss of rate for both the search presented in this work and for that of our earlier Ref.~\cite{Acanfora:2023gzr}.}

As already noted in Ref.~\cite{Acanfora:2023gzr}, the \BelleII geometry does not allow to fully take advantage of the photon-fusion production because of the limited acceptance to forward electron and positrons. The current angular coverage of Belle~II can be written in terms of the electron (positron) pseudorapidity in the center of mass frame as
\begin{equation}
\text{Belle~II}:\quad    |\eta^*_{e^\pm}|=\left|\log( \tan ({\theta^*_{e^\pm} \over 2}))\right|\leq 1.64\ , \label{eq:BelleIIopen}
\end{equation}
and corresponds to electron and positron with a \emph{minimal} angle of $22\degree$ with respect to the beam axis. 

Since the limited forward acceptance of Belle~II cuts away a large fraction  of the fusion production cross-section, it is interesting to explore how the reach of diphoton resonances could be improved by a hypothetical  experiment with a wider angular coverage at an $e^+e^-$ facility. The result for this hypothetical detector, which we call Belle-fwd, are presented in green in Fig.~\ref{fig:money} assuming the same luminosity of Belle~II. In order to fix a benchmark, we imagine a pseudo-rapidity coverage in the center of mass frame for both positrons and electrons of
\begin{equation}
\text{Belle-fwd}:\quad |\eta^*_{e^\pm}|=\left|\log( \tan ({\theta^*_{e^\pm} \over 2}))\right|\leq 5\ ,\label{eq:bellefwd} 
\end{equation}
which corresponds to a minimal angle with respect to the beam axis of $0.76\degree$. For reference, this roughly corresponds to the forward acceptance of the ATLAS calorimeter.
 
The possibility of extending the forward coverage at Belle~II is limited by the presence of collimating magnets around the beam line which do not allow this region to be instrumented with detectors. The magnets are  required to reach the expected integrated luminosity of the Belle~II experiment~\cite{Belle-II:2010dht}. Thanks to the impressive boost in the signal rate for an extended forward acceptance, even with reduced luminosity it is still possible to get an improvement in the reach compared to the Belle~II experiment. This of course will also reduce the challenge of squeezing the beam to reach the luminosity targeted at Belle~II. A more detailed illustration of how the reach in this channel depends on the forward acceptance and the luminosity is shown in Fig.~\ref{fig:eta_lumi_contours} and discussed in Sec.~\ref{sec:forwardfactory}.  

We conclude in Sec.~\ref{sec:conclusion} leaving the technical details of our study for a series of appendices. In App.~\ref{sec:MC} we discuss how we overcome the challenges of generating a smooth SM background prediction in our binned 4D likelihood analysis. In particular, we discuss the different possible statistical treatments of the empty bins in the MC, which result in the uncertainty in the expected reach in Fig.~\ref{fig:money}. In App.~\ref{sec:likelihood} we further detail the features of our likelihood. In App.~\ref{app:EPA} we quantify the range of applicability of the well-known effective photon approximation (EPA)~\cite{Fermi:1925fq,Williams:1934ad,Budnev:1975poe,Frixione:1993yw} that is often advocated to carry out simplified cross-section computations and to develop intuition on the signal kinematics. App.~\ref{app:lifetime} discusses the event by event reweighting we performed to account for the ALP lifetime.

\section{Expected sensitivity at Belle~II  \label{sec:Belle2} }

In our signal we deal with $e^{\pm}$ and $\gamma$ detected in  Belle~II.  According to \BelleII detector performance~\cite{Belle-II:2018jsg} these objects are well reconstructed if they satisfy the following energy and angular requirements:
\begin{equation}
\label{eq:acceptance}
E^*>0.25\GeV,\quad\, \theta^*\in [22,158]\degree\,,
\end{equation}
where the starred quantities are measured in the collision center of mass frame.  In addition, as we want to discuss a resonance in the spectrum of resolved photons, we require 
\begin{equation}\label{eq:48-mrad}
\Delta \theta_{\gamma\gamma} > 0.048\quad  {\text{ or }} \quad\Delta\phi_{\gamma\gamma} > 0.048\ ,
\end{equation}
where $\Delta \theta_{\gamma\gamma}$ and $\Delta\phi_{\gamma\gamma}$ indicate the polar and azimuthal angular differences between the two resonant photons. This requirement ensures that the two photons in the signal final state are reconstructed as separated photons at Belle~II~\cite{Belle-II:2018jsg}. 

The photons need also to be separated from the electron to give rise to a clean final state, thus we require for all $e^{\pm}$-photon  pairs
\begin{equation}
\Delta \theta_{\gamma e^{\pm}} > 0.048 \quad  {\text{ or }} \quad \Delta\phi_{\gamma e^{\pm}} > 0.048\ .
\label{eq:48-mrad-e}
\end{equation}
The angular separation and energy requirements of \cref{eq:acceptance,eq:48-mrad,eq:48-mrad-e} guarantee the convergence of the perturbative expansion so that both our signal and background samples 
are generated with \madgraph~v2.7.3~\cite{Alwall:2014hca} at leading order in perturbation theory. 

\subsection{Signal vs Background \label{sec:signal}}
Two possible ALP production mechanisms at a lepton collider are
\begin{eqnarray}
e^+e^- &\to& \gamma_{\text{vis}} a \, , \label{eq:strahlung}\\
e^+ e^- &\to& e^+_{\text{vis}} e^-_{\text{vis}} a \, ,\label{eq:fusion-visible}
\end{eqnarray}
with $a\to \gamma\gamma$. We have denoted  by the subscript $vis$ the final states that are in the  acceptance of the \BelleII detector~\cite{Belle-II:2018jsg} given in \eq{eq:acceptance}. Depending on the lifetime of the ALP, the production mechanisms above can give rise to either a visible resonant diphoton pair $a\to(\gamma\gamma)_{\text{res}}$,  the processes \eq{eq:strahlung} and \eq{eq:fusion-visible}, or to missing energy and momentum $a\to\text{invisible}$.

\begin{figure}[h!]
\centering
\includegraphics[width=0.45\linewidth]{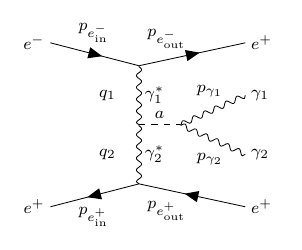}\hfill
\caption{Feynman diagram for the $e^+ e^- \to e^+ e^- a$ production.}
\label{fig:ee_eea_tchan_diagram}
\end{figure}

The signal in \eq{eq:strahlung} is the well studied ``ALP-strahlung'' production (see Ref.~\cite{Dolan:2017osp}), while the process in \eq{eq:fusion-visible} gives rise to two different processes: $i$)~the ``ALP-Dalitz'' process, given by the ALP-strahlung \eq{eq:strahlung} with a photon conversion into $e^{+}e^{-}$; $ii$) the ``photon-fusion'' into ALP, given by  a photon line exchanged in the $t$-channel that radiates the ALP depicted in \cref{fig:ee_eea_tchan_diagram}. Similarly to what we found in \cite{Acanfora:2023gzr} for the invisible ALP decay, the sensitivity for an ALP decaying into a diphoton pair at Belle~II is dominated by the photon-fusion production. This dominance would be further enhanced in an experiment with extended forward detector coverage.

For each putative $m_{a}$ we require the two final state  photons to have an invariant mass in a range 
\begin{equation}
m_{\gamma \gamma} - m_a \in [-3,1.5]\cdot \sigma_{m_a} \,,\label{eq:bump_hunt}
\end{equation} where $\sigma_{m_a}$ denotes the expected resolution on a di-photon resonance~\cite{Belle-II:2018jsg}, which is dominated by the uncertainty on the photon energy depositions in the ECAL. This requirement removes large fractions of the  backgrounds for all $m_a$ of interest and entirely kills the Dalitz contribution to the signal, resulting in a very small loss of signal rate. 

The most relevant background to the photon fusion production in Eq.~\eqref{eq:fusion-visible} is given by radiative corrections to  Bhabha scattering
\begin{equation}
e^+e^- \to e^+e^- \gamma \gamma\, \label{eq:bhabha}.
\end{equation}
A large fraction of the cross-section for this process can be understood qualitatively as $e^+e^- \to e^+e^-$ dressed by two $e^\pm \to e^\pm \gamma$ splittings. As a consequence of the soft and collinear enhancement of the splittings, a distinguishing feature of this background process is the tendency of each  photon to fly collinear to the fermion that has emitted it, as well as being preferred to be soft compared to the parent $e^\pm$.  
As a further consequence of the collinear preference of the photons emerging from Bhabha scattering, we observe that a large fraction of the background events  tends to have photons flying in opposite directions. This is expected when the two photons are arising from different fermionic lines in the hard scattering. 

\begin{figure*}[h!]
\includegraphics[width=0.45 \textwidth]{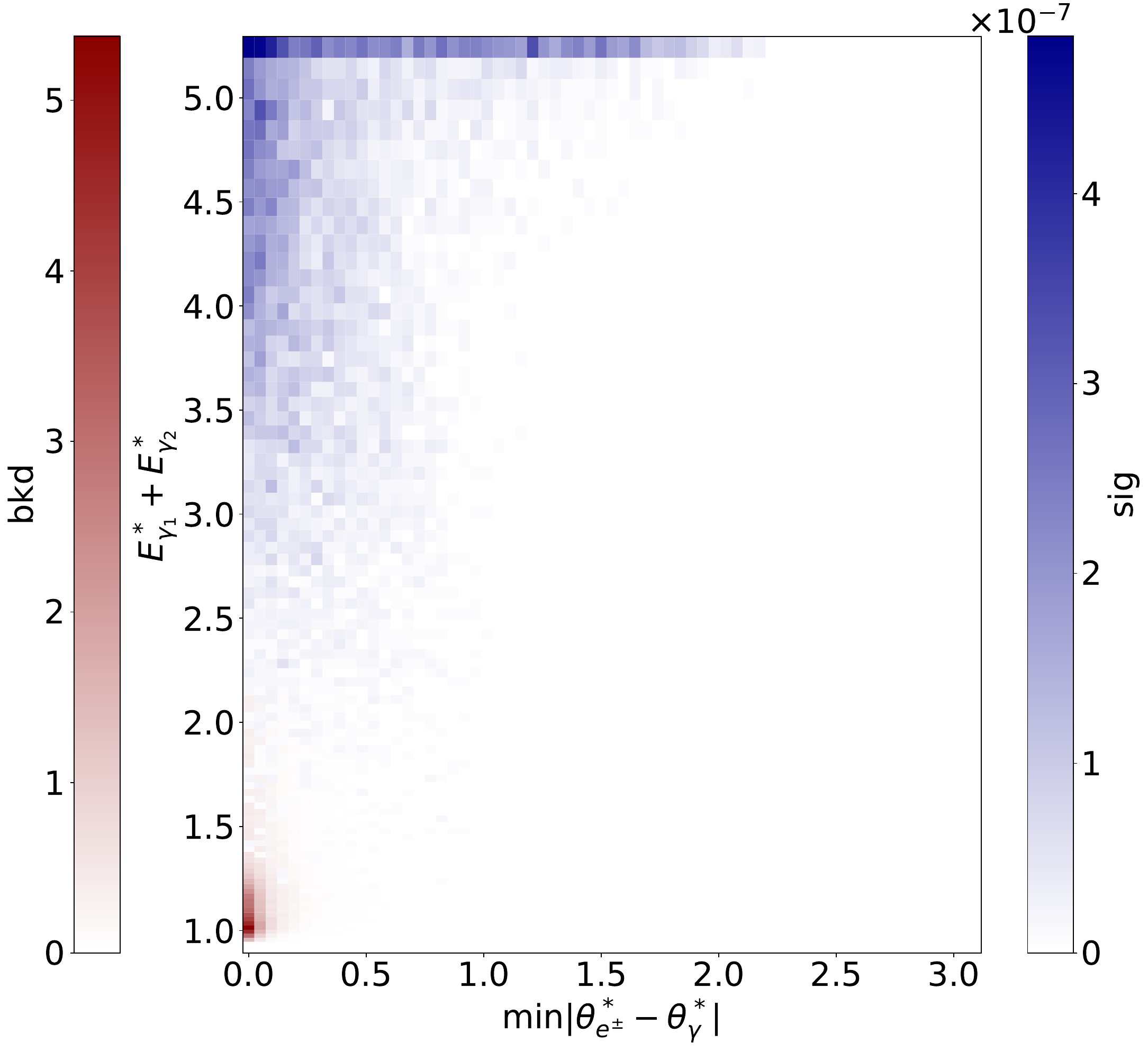}\hfill
\includegraphics[width=0.464 \textwidth]{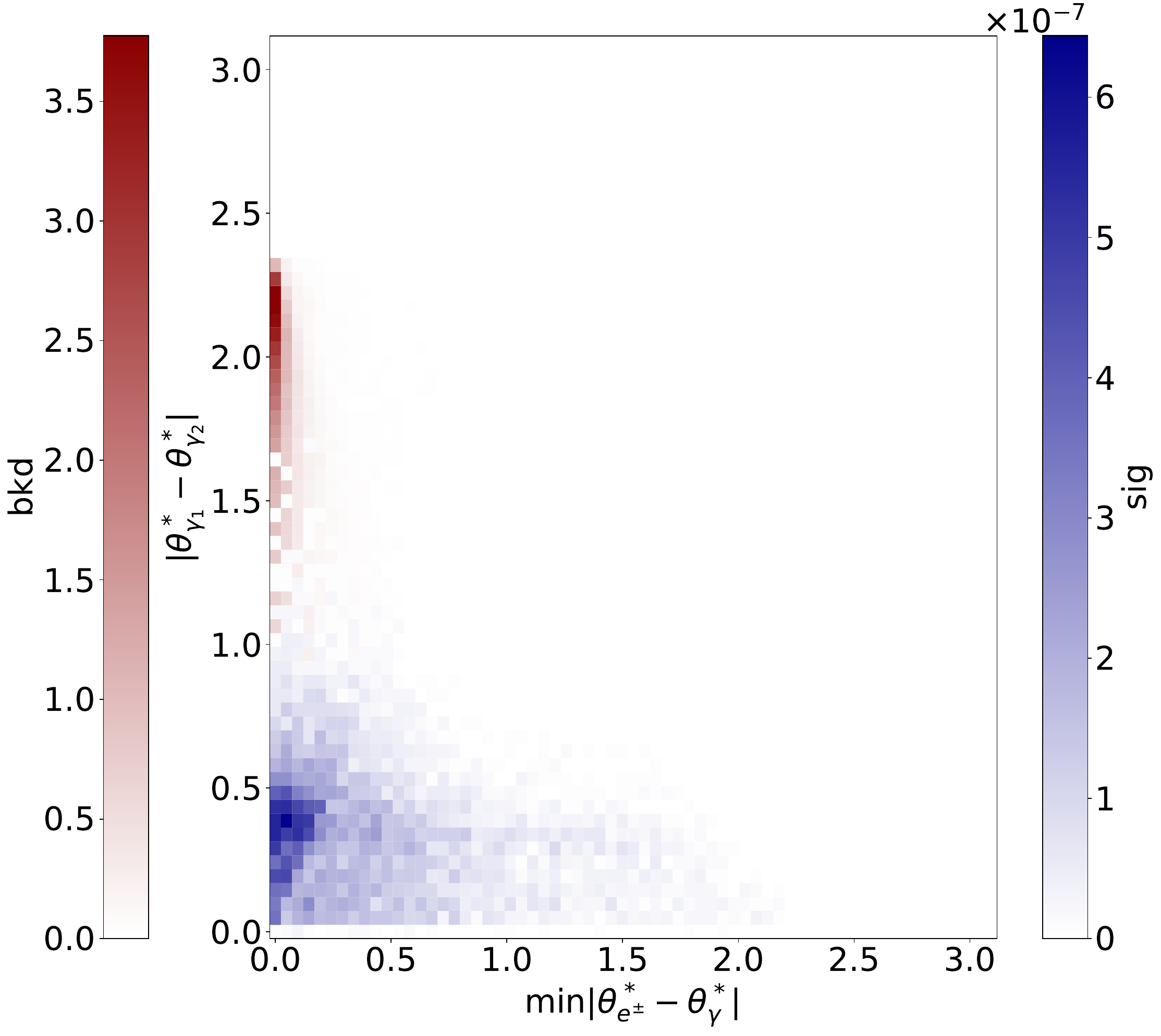}
\caption{Distributions of signal events for the ALP production in photon fusion in Eq.~\eqref{eq:fusion-visible} (in {\color{navy}\bf blue}) and background events for the radiative Bhabha in Eq.~\eqref{eq:bhabha} (in {\color{red}\bf red}). A fixed diphoton invariant mass of $m_{\gamma\gamma}=1\text{ GeV}$ within the range of Eq.~\eqref{eq:bump_hunt} is imposed, as well as the acceptance and isolation cuts at Belle~II given in \cref{eq:acceptance,eq:48-mrad,eq:48-mrad-e}. {\bf Left:} distributions as a function of the minimal angle between the positron/electron direction and the photon and the total energy of the photon pair, {\bf Right:} distributions as a function of the minimal angle between the positron/electron direction and the photon and the angular separation between the two photons.}
\label{fig:tplus-tminus-splus-sminus}
\end{figure*}
In Fig.~\ref{fig:tplus-tminus-splus-sminus} the distributions of signal and background at fixed invariant mass of 1 GeV are shown. As discussed above, the background collinear enhancement will favor a small angle between the photon and the corresponding fermionic line (either the electron or the positron). As a consequence, in both panels in Fig.~\ref{fig:tplus-tminus-splus-sminus} we can see how the background event distribution favors a small \emph{minimal} angle between the photon and electron (or positron) in the final state. 

Moreover, for a small enough  diphoton invariant mass, the signal events 
passing the Belle~II acceptance cut typically feature a boosted ALP resonance (as already shown in Ref.~\cite{Acanfora:2023gzr} but see also Fig.~\ref{fig:mean_median_pt_1_8_gev} in App.~\ref{app:EPA}), which results in an energetic photon pair with relatively small opening angle between the photons. Conversely, the enhancement of the splitting function for soft radiation prefers the photons energies to be small and the angle between them could easily be large to make the required invariant mass if for instance they are split in opposite directions. The separation between signal and background becomes less sharp at higher invariant masses, when the boost of the signal resonance diminishes. This feature qualitatively explains the degradation of the sensitivity at Belle~II apparent in Fig.~\ref{fig:money}. 

Before leaving this section, we comment on further backgrounds beyond the one in Eq.~\eqref{eq:bhabha}. One possibility is the rare mis-reconstruction of hadrons as electrons and fake photons, e.g. from mis-identification of neutral mesons, e.g. a $\pi^0$ detected as a photon. The efficiency-vs-purity performance of the \BelleII experiment on particle-ID is a subject of active development (see e.g. Ref.~\cite{Sebastian:1821}). A realistic estimate of these backgrounds is challenging, especially when multiple hadrons arise from a copious two-body hard scatterings such as $e^{+}e^{-}\to q\bar{q}$. The final background rate depends mainly on the detector operation, e.g. isolation requirements, and should be performed by the experimental collaboration.

\subsection{Sensitivity}\label{sec:sensitivity}

For an optimal signal vs background separation we pursue a full four dimensional likelihood, which can take full advantage of their different kinematic structure. In order to efficiently tessellate the four dimensional phase space we define, in analogy with the usual Mandelstam variables, the Lorentz invariants $t_\pm$ and $s_\pm$ as
\begin{equation}
\begin{split}
\tplus &=\left( p_{ e_{in}^{+}  } - p_{e_{out}^{+}} \right)^{2} \,,\\ 
\tminus &=\left( p_{ e_{in}^{-}  } - p_{e_{out}^{-}} \right)^{2} \,,\\   
\sminus &=\left( p_{ e_{in}^{+}  } + p_{ e_{in}^{-}  } - p_{e_{out}^{+}} \right)^{2}\,,\\   
\splus &=\left( p_{ e_{in}^{+}  } + p_{ e_{in}^{-}  } - p_{e_{out}^{-}} \right)^{2} \, ,    
\end{split}
\label{eq:s12t12}
\end{equation}
where the positron, electron, and photon momenta are defined in Fig.~\ref{fig:ee_eea_tchan_diagram}. A conversion between the Lorentz invariant variables and the measured energy and angles of the positron, electron, and photons can be easily derived. Neglecting the electron mass, we can write 
\begin{equation}
t_{\pm}\simeq -\sqrt{s} E_{e_{out}^\pm}\left(1 \pm \cos\theta_{e_{out}^{\pm}}\right)\quad ,\quad s_{\pm}\simeq\sqrt{s} \left(\sqrt{s}-2 E_{e_{out}^{\pm}}\right)\ ,\label{eq:definitions}
\end{equation}
where $\theta_{e_{out}^{\pm}}$ is the polar angle of the positron/electron with respect to the beam axis. The above expressions show that $t_{\pm}\to 0$ controls the collinear singularity of the photon fusion production at $\theta_{e_{out}^{\pm}}\to0$. Using energy conservation, the sum of the other two Mandelstam variables can be written in terms of the energy of the resonance $s_{+}+s_{-}\simeq 2E_a\sqrt{s}$, while their difference gives the asymmetry between the positron and the electron energy.

Starting from the energy and polar angle resolution of \BelleII~\cite{Belle-II:2018jsg} we derive approximate resolutions for the measurement of $t_\pm$ and $s_\pm$. Each of these quantities contains energies and directions of photons and electrons. The energies are the least well measured, so we can take 
\begin{equation}
\frac{\delta \mu_{i}}{\mu_{i}}\simeq 2 \frac{\delta E}{E}\simeq 4\%\quad\text{for}\quad  \mu_{i}=\{\splus, \sminus, \tplus, \tminus\}\ .\label{eq:BelleIIres}
\end{equation}

Based on the predicted differential rates for the SM background and for the signal, we construct a log-likelihood for 50~ab$^{-1}$ integrated luminosity at \BelleII
\begin{equation}
\Lambda=-2 \sum_{i,j} \ln \frac{L(S_{i,j},B_{i,j})}{L(0,B_{i,j})} \label{eq:likelihood} \,,
\end{equation}
where $i$ and $j$ run on the bins of the 4D space spanned by $\splus$, $\sminus$, $\tplus$,   $\tminus$ and $S$ and $B$ indicate respectively the expected number of signal or background events in each bin. In \eq{eq:likelihood} $L(S_{i,j},B_{i,j})$ is the Poisson factor computed in each bin defined as follows
\begin{equation}
L(S,B)=\frac{\left(S+B\right)^{B}}{B !} e^{-(S+B)}\,.
\end{equation}
The  sensitivity shown in Fig.~\ref{fig:money} corresponds to 95\% C.L. and it is obtained by requiring $\Lambda<4$. 

Given the huge hierarchy between the signal rate and the background rate, our sensitivity benefits from regions of phase space where the background is strongly suppressed. To best exploit these features, we perform a full 4D likelihood analysis. This likelihood is challenging to construct from the viewpoint of the MC generation, but the improvement in the sensitivity is worth the effort.\footnote{We note in passing that the expected sensitivity for a 3D likelihood obtained by collapsing our 4D one in any of the four directions is reduced by roughly one order of magnitude. This loss of sensitivity confirms that the full phase space information is needed to distinguish the photon fusion process from the background.}  For computational reasons we then tessellate the 4D phase space with different bin sizes depending on the ALP mass. Specifically for $m_a< 1\text{ GeV}$ we take $\delta \mu_{i}/\mu_{i}=15\%$, for $1\text{ GeV}\leq m_a< 3\text{ GeV}$ we take $\delta \mu_{i}/\mu_{i}=8\%$ and for $m_a\geq3\text{ GeV}$ we take the realistic Belle~II resolution estimated in Eq.~\eqref{eq:BelleIIres}. In this sense, the expected reach of Fig.~\ref{fig:money} at low masses should be considered conservative and could be improved by reducing the bin size further. 

Even with coarse grained bins, our expected reach has a residual dependence on the statistical treatment of the phase space cells with zero expected background within our MC uncertainty. Evaluating this uncertainty ultimately results in the uncertainty band in the expected reach shown in Fig.~\ref{fig:money}. A detailed description of our methods to generate a smooth background sample in the full 4D phase space and assigning a conservative uncertainty to it is given in Appendix~\ref{sec:MC}.  

We also checked the robustness of our result against possible systematic uncertainties. These are particularly dangerous as one approaches a high statistics regime and small couplings can be potentially probed. To compute the loss of sensitivity due to systematics we computed the log-likelihood keeping only bins for which the expected $S/B$ was larger than a certain threshold $\epsilon_{
\rm{sys}}$. For $\epsilon_{\rm{sys}}<10\%$ the expected sensitivity does not change appreciably, especially at low masses. This is because the likelihood is dominated by bins with very suppressed background yield, as discussed in detail in Appendix~\ref{sec:likelihood}.

The variable $\tplus, (\tminus)$ in Eq.~\eqref{eq:s12t12} is related to the degree of collinearity  of the initial and final $e^{+}, (e^-)$, which in turn can be seen as the collinearity of the virtual photons and the initial $e^{+}, (e^-)$. These quantities are typical markers of $t$-channel scattering, which dominates the  photon-fusion production of the ALP. As it can be appreciated from Fig.~\ref{fig:likelihood} in Appendix~\ref{sec:likelihood}, the Belle~II acceptance cuts off this singularity reducing the signal acceptance. In the next section we show how a hypothetical experiment with extended forward acceptance would allow instead to explore kinematic configurations where both background and signal are closer to the collinear singularity and can be effectively distinguished by exploiting the full four dimensional phase space information.

\begin{figure}
\centering
\includegraphics[width=0.45\linewidth]{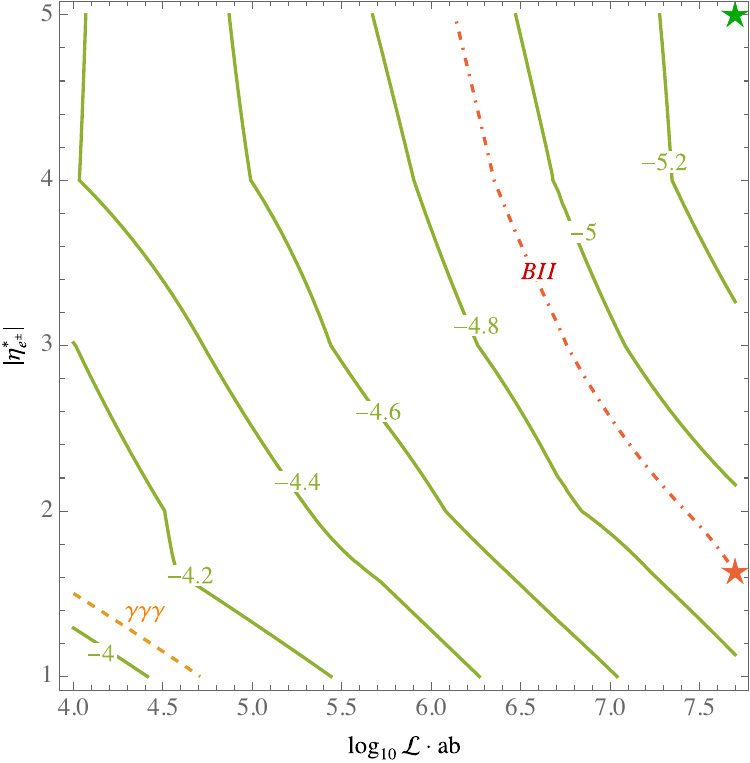}
\caption{The {\color{green}\bf green contours} show the expected reach for $\log_{10}(g_{a \gamma \gamma})$ in the $e^+ e^- \to e^+ e^- (a \to \gamma \gamma)$ channel as a function of the luminosity and the final $e^\pm$ acceptance of a hypothetical lepton collider at $\sqrt{s}=10.58~\GeV$, given $m_a=3$ GeV. The {\color{red}\bf red} and  {\color{green}\bf green} stars are the benchmarks for the Belle II and Belle-fwd reach in Fig.~\eqref{fig:money}. 
As a reference we show as a  {\color{orange}\bf orange} dashed line the Belle II reach for the $e^+ e^- \to \gamma (\gamma \gamma)_{\text{res}}$ channel. The red dashed line is the isoline of the value of the Belle II reach.
}
\label{fig:eta_lumi_contours}
\end{figure}

\section{Expected sensitivity of a forward flavor factory}\label{sec:forwardfactory}

In this section we entertain the possibility that a future experiment  with extended forward coverage will be available at an $e^+e^-$ facility similar to SuperKEKB. Given that the expected luminosity of such a machine is likely to be correlated with its forward instrumentation, we want to study how the expected reach of the search discussed in Sec.~\ref{sec:Belle2} depends on the forward acceptance, $\eta^*_{e^\pm}$, and the luminosity, $\mathcal{L}$. We do not make a concrete technical proposal here on how to construct and operate this extended detector. We remark however that this is an active area of research for future collider detectors studies \cite{Accettura:2023ked,Ruhdorfer:2023uea}.

Our results are shown in Fig.~\ref{fig:eta_lumi_contours} for benchmark ALP mass of 3 GeV and $\sqrt{s}=10.58 \GeV$, where the red and the green stars correspond to the Belle~II reach and the Belle-fwd reach of Fig.~\ref{fig:money}. The reach in $g_{a\gamma\gamma}$ defined in Eq.~\eqref{eq:ALPlagrangian} decreases roughly like $\mathcal{L}^{1/4}$ as expected for a search that is not free from background. Conversely, increasing the forward acceptance leads to a steep rise of the sensitivity which decreases only at high forward acceptances. We find  that, thanks to  the impressive gain in reach showed in Fig.~\ref{fig:money}, even an experiment collecting 100 times less luminosity than Belle II, profiting of an extended forward acceptance down to $\eta_{e^\pm}^*<5$, could lead to a substantial improvement in the reach compared to Belle~II. 



\section{Conclusion}~\label{sec:conclusion}
In this work we developed a new search for diphoton resonances at Belle~II based on the photon fusion production in Eq.~\eqref{eq:signal-intro}. By exploiting a full tessellation of the four dimensional phase space of this $2\to3$ process we performed a detailed study of how this can be separated from the background given by the radiative Bhabha process in Eq.~\eqref{eq:bhabha}. The result of this analysis is summarized in Fig.~\ref{fig:money} which shows how this channel can lead to a substantial improvement in the reach compared to the ``ALP-strahlung'' production studied in Ref.~\cite{Dolan:2017osp} at least for ALP masses below 6 GeV. Above 6 GeV the Belle II reach quickly becomes less sensitive than existing constraints from ultra-peripheral heavy ions collisions~\cite{Knapen:2017ebd,CMS:2018erd}. Along the way, we illustrate in Appendix~\ref{sec:MC} a method to associate a conservative uncertainty to the signal reach in analyses plagued by challenges in the MC sampling like ours. 

In addition, we explored how the expected reach for a diphoton resonance in the $e^+e^-+\gamma\gamma$ channel can be further improved at a hypothetical experiment with extended forward coverage compared to Belle~II. The results of Fig.~\ref{fig:money} together with the ones in Fig.~\ref{fig:eta_lumi_contours} show that with an extended forward coverage down to $|\eta_{e^\pm}^*|<5$ the expected Belle~II reach could be improved even with 100 time less luminosity.

\section*{Acknowledgments}
We thank Torben Ferber, Enrico Graziani, Antonio Passeri and Laura Zani for useful discussions about Belle~II. RF and DR are supported in part by the European Union - Next
Generation EU through the MUR PRIN2022 Grant n.202289JEW4. 

\appendix

\section{Generation of the background sample for the 4D likelihood analysis\label{sec:MC}}

The generation of 4-fold differential cross-section requires intense computational resources. A relatively gross-grained sampling of each dimension with about 100 bins leads to a rather large set of $10^8$ bins in a single histogram. Obtaining the value of the cross-section with precision $1/p$ in each bin requires $p^2$ samplings of the differential cross-section, which, even for leading order matrix elements, takes some substantial time. 

Significant speed-up of the generation of the information necessary to characterize our background in a 4D phase-space can be obtained  i) parallelizing the generation of multiple MC samples, ii) using a clever re-weighting of the events, iii) applying suitable biases to the samples, and iv) combining  the information from the several runs.

In our calculation we have applied a bias to the differential cross-section as to overpopulate tails at large $\splus$ and large $\sminus$, and to  increase the population of the area of phase-space around $\tplus\simeq\tminus\simeq 0$. These regions are most difficult to describe precisely with \madgraph~ both because they are tails 
and because they are rather narrow region with respect to  the whole phase-space of the process.
{In addition, $\tplus\simeq\tminus\simeq 0$ is a boundary of the physical phase-space, which poses extra difficulties.}
The accurate description of these regions of phase-space is of utmost importance for our results as they give a large discrimination of signal over background within our likelihood in Eq.~\eqref{eq:likelihood}. 

Particularly challenging is the generation of background events for  $m_a\leq 1$ GeV at Belle II, because the kinematics of the signal is such that the background suppression in specific regions of phase space is larger than for higher masses. In what follows we describe our MC strategy in this light mass range and mention the simplification at higher masses at the end. 

For  $m_a\leq 1$ GeV, we produce ${\cal O}(10)$   MC samples with \madgraph~ with bias and cuts aimed towards several different combinations of small and large $\splus+\sminus$ and similarly for $\tplus,\tminus$. From each sample we obtain a weight, that corresponds to the cross-section for each given cell of the 4D phase-space. We denote this weight as $w_{c,s}$, where $c$ denotes the cell of the phase-space and $s$ the sample we are dealing with. Each such weight can be endowed of an {\it estimated} { uncertainty}, which  originates from the finiteness of the MC samples that we generated. We estimate an uncertainty $\delta w_{c,s} = w_{c,s}/\sqrt{n_{c,s}}$, where $n_{c,s}$ is the number of weighted MC events in the cell $c$ from the sample $s$. 

We then combine the weights of a given cell $c$ as if they were the results of different experiments into a combined value
\begin{equation}
w_{c,C} = \frac{\sum_s w_{c,s} \cdot \xi_{c,s}}{\sum_s \xi_{c,s}}\, , \label{eq:central-gauss}
\end{equation}
where $\xi_{c,s}=\delta w_{c,s}^{-2}$, as in the usual combination of two or several measurements described by an underlying Gaussian probability density function. Linear uncertainty propagation is applied to this formula, finding the standard result for the uncertainty on the combined value
\begin{equation}
\delta w_{c,C} = \left( \sum_{s} \delta w_{c,s}^{-2}   \right)^{-1/2}\,.  \label{eq:std-gauss}
\end{equation} 

The samples $s$ included in the combinations are chosen so that only measurements that are {\it compatible} with each other are used. In particular, we combine only samples that have non-zero weight on the cell $c$ and discard the samples that returned zero weight, either because $c$ was entirely out of the region of phase-space of the MC run that produced sample $s$, or because the bias in that run was put to highlight other tails of the phase-space. The runs with zero weight are discarded on the ground that they correspond to incompatible measurements, and therefore cannot be combined with the other group. In fact, the information from the samples returning zero weight would imply zero cross-section on that cell, which is clearly incompatible with the information from a dedicated sample that has found non-zero cross-section in that cell. 

When a cell is not populated by any of our background samples simulations, we conservatively assume that the cross-section in that cell is non-zero and that each sample has given us an upper-bound on the cross-section in that cell. Each sample contributes an upper-bound on the weight $w_{c,s}<\Delta w_s$, where $\Delta w_s$ is the minimum weight 
in  the sample $s$.   The combined upper-bound then is  
\begin{equation}
w_{c,C}< \Delta w_{c,C}\quad\text{ for }\quad \Delta w_{c,C} = \left( \sum_s \Delta w_s^{-2} \right)^{-1/2}\,.\label{eq:upperbound}
\end{equation}

From the above procedure we obtain a combined histogram in 4D that has either a measurement or an upper bound on the cross-section in each cell of phase-space. 
 In the following we describe a method to assess the impact of the choice in the treatment of these cells for which we have obtained an upper bound.  

In order to give an estimate of the error on the bounds on $\gagg$ due to the finiteness of our MC samples, we produce replicas of our combined histogram. In each replica we fluctuate the number of expected events in the cell $c$ based on the underlying uncertainty on the combined weight $\delta w_{c,C}$. For the bins with non-zero background events the number of the expected events follows a truncated normal distribution where the gaussian mean and standard deviation are taken from Eq.~\eqref{eq:central-gauss} and Eq.~\eqref{eq:std-gauss}. 

For the empty bins we only have the upper bound in Eq.~\eqref{eq:upperbound}, which leaves us the freedom of picking the probability distribution  in the bin. Variations of the underlying probability distribution for the bins gives the uncertainty in the expected reach in Fig.~\eqref{fig:money}. We identify three scenarios:
\begin{itemize}
    \item The weakest constraints come from assuming that the  distribution is a Dirac $\delta$ function centered at the upper bound in Eq.~\eqref{eq:upperbound}. This gives the maximal number of expected background events compatible with having observed zero events in our Monte Carlo.
    \item The strongest constraints are obtained assuming a  discrete Poisson distribution with mean set at the upper bound in Eq.~\eqref{eq:upperbound}. The discrete nature of this distribution  gives a sizable probability  to have zero expected background, thus boosting the likelihood in Eq.~\eqref{eq:likelihood}.
    \item Intermediate bounds between these two options are obtained by picking a continuous uniform probability distribution from zero to the upper bound in Eq.~\eqref{eq:upperbound}. The central line in the bands in Fig.~\eqref{fig:money} corresponds to this choice, which we consider as the most reasonable estimate of the attainable constraints.
\end{itemize}

The background generation for $m_a\geq 1\text{ GeV}$ is analogous to the one explained above, with the technical simplification that a single MC sample was sufficient to obtain a smooth background sample. Running multiple samples would have allowed us to reduce the bin size of the likelihood for $1\text{ GeV}\leq m_a<3\text{ GeV}$ 
and in general reduce the uncertainty in our projected sensitivity for all the ALP masses. The projected sensitivity and its uncertainty shown in Fig.~\ref{fig:money} should then be taken as a first conservative assessment of the discriminating power of the $e^+e^-\to e^+e^- (\gamma\gamma)_{\text{res}}$ channel. 

\begin{figure*}[t!]
\centering
\includegraphics[width=1\linewidth]{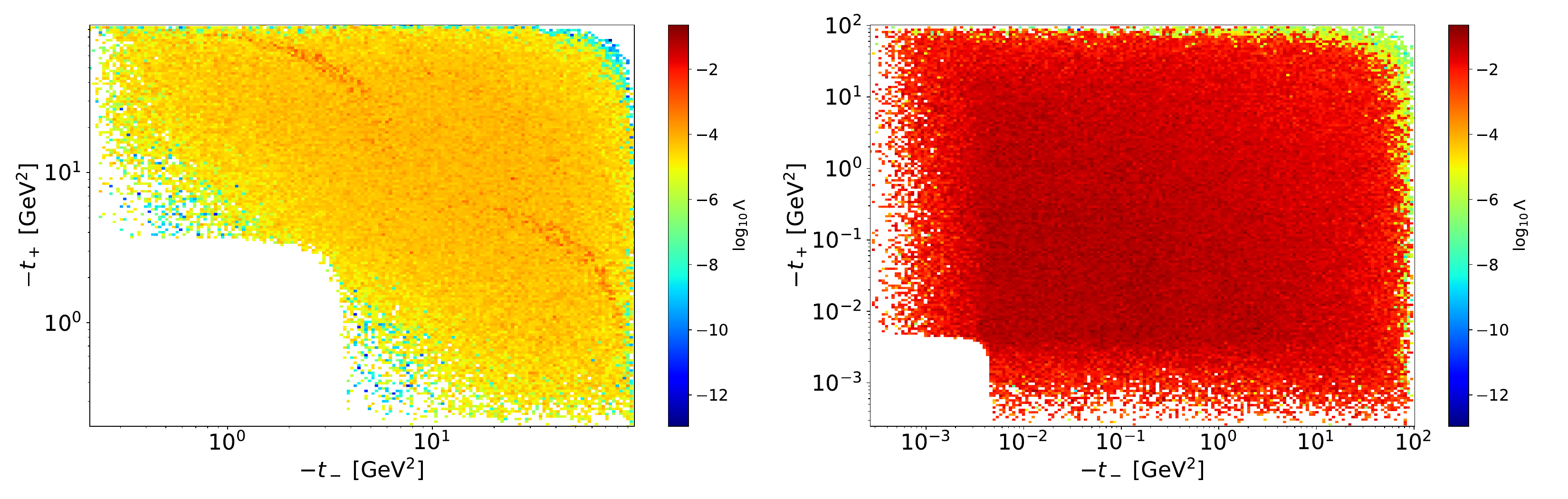}
\caption{Distribution of log-likelihood defined in Eq.~\eqref{eq:likelihood} in the $(t_-,t_+)$ plane for $m_a=3\text{ GeV}$. The 4D likelihood is maximized in the $(s_-,s_+)$ plane to be projected on a 2D plane. The signal strength is fixed to the expected reach of Belle-fwd with 50 $\text{ab}^{-1}$ of data which is $g_{a \gamma \gamma}=5.5 \cdot 10^{-6}$ GeV$^{-1}$ as shown in Fig.~\ref{fig:money}. The white region is kinematically forbidden because of the detector acceptance. \textbf{Left} Belle II likelihood with acceptance defined in Eq.~\eqref{eq:BelleIIopen} \textbf{Right} Belle-fwd likelihood with acceptance defined in Eq.~\eqref{eq:bellefwd}.}
\label{fig:likelihood}
\end{figure*}

\section{Features of the 4D likelihood\label{sec:likelihood}}

\begin{figure}
\centering
\includegraphics[width=1\linewidth]{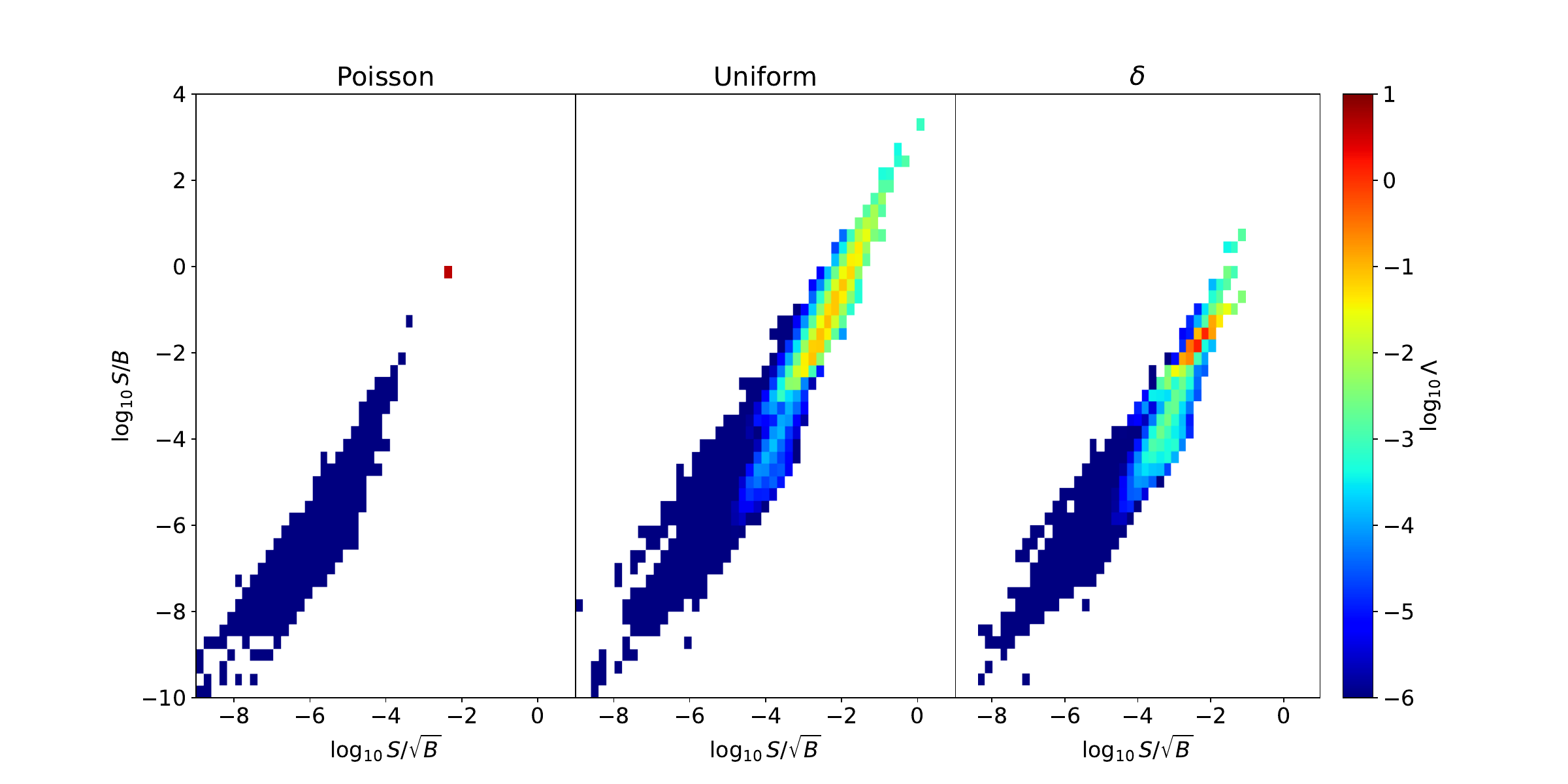}
\caption{Likelihood distributions for $m_a=3\text{ GeV}$ in the plane $(S/\sqrt{B}, S/B)$ where $S$ is the expected number of signal events and $B$ the expected number of background events. The three panels correspond to the three different assumption on the distribution of the background events in the empty bins as illustrated in Appendix~\ref{sec:MC}. {\bf Left:} The events in the empty bins are distributed as a Poisson distribution with mean set to the upper bound of the cross-section derived in our MC. The likelihood correspond to the best expected sensitivity in Fig.~\ref{fig:money} which is $g_{a\gamma\gamma}=5.49 \cdot 10^{-6} \text{ GeV}^{-1}$ {\bf Center:} The events on the empty bins are distributed as a uniform distribution with maximum given by the upper bound derived in our MC. This treatment correspond to the red line in Fig.~\ref{fig:money} which gives $g_{a\gamma\gamma}=2.88 \cdot 10^{-5} \text{ GeV}^{-1}$. {\bf Right:} The events on the empty bins are distributed as a delta-function centered at the upper bound derived in our MC. This assumption gives the most conservative expected reach in Fig.~\ref{fig:money} which is $g_{a\gamma\gamma}=5.89 \cdot 10^{-5} \text{ GeV}^{-1}$.}
\label{fig:ll_syst_triplet}
\end{figure}
We now comment on the features of the 4D likelihood which result in the sensitivity at Belle~II and at Belle-fwd in Fig.~\ref{fig:money}.  Firstly, we show in Fig.~\ref{fig:likelihood} the distribution of our likelihood as defined in Eq.~\eqref{eq:likelihood} in the $(t_{-}, t_{+})$ plane. For every $t_{-}, t_{+}$ pair, we pick the maximal likelihood value in the $(s_-,s_+)$ plane. In the left plot we see the distribution at Belle~II while in the right one at a hypothetical experiment  with extended forward coverage at an $e^+e^-$ facility and the same luminosity of Belle~II. 

\begin{figure}
\centering
\includegraphics[width=0.46\linewidth]{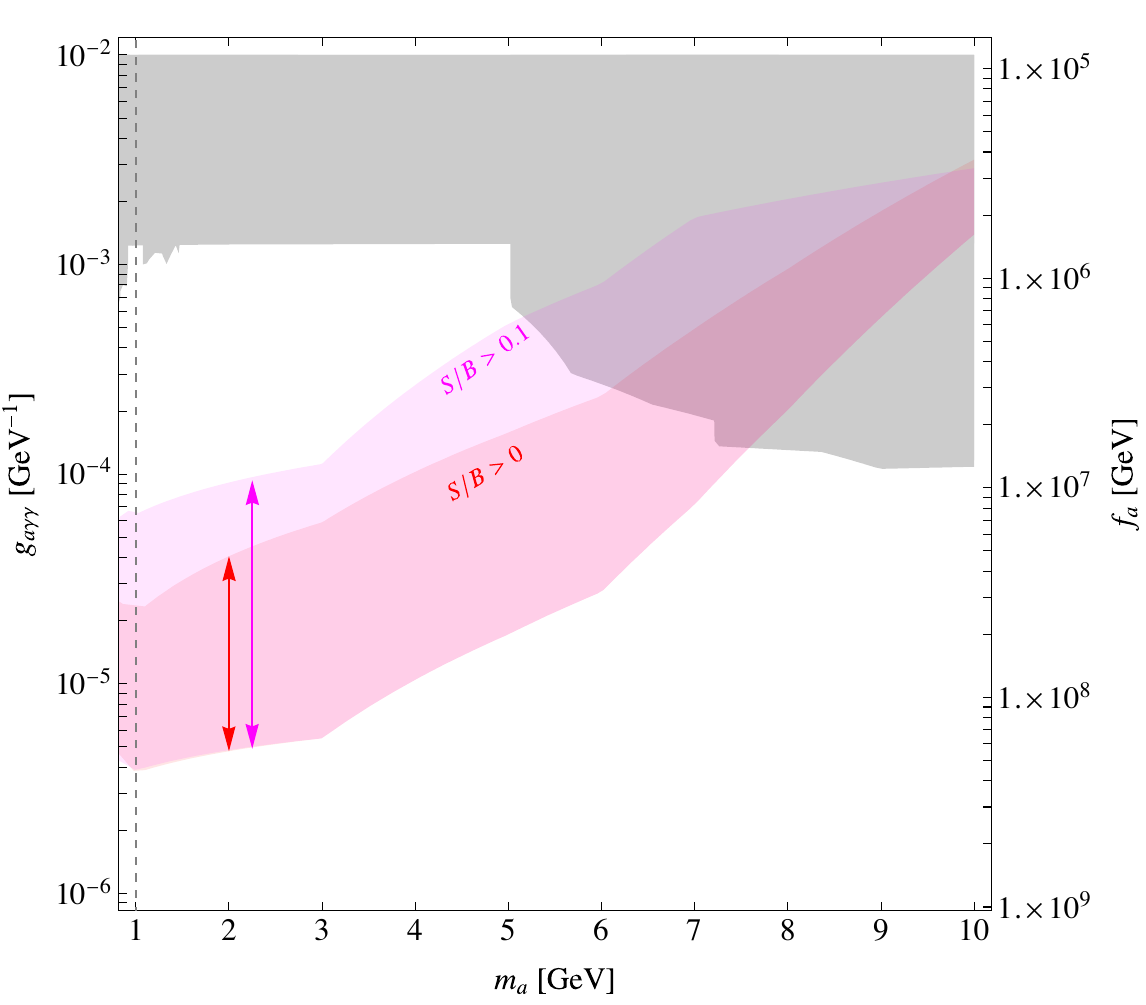}\hfill
\includegraphics[width=0.46\linewidth]{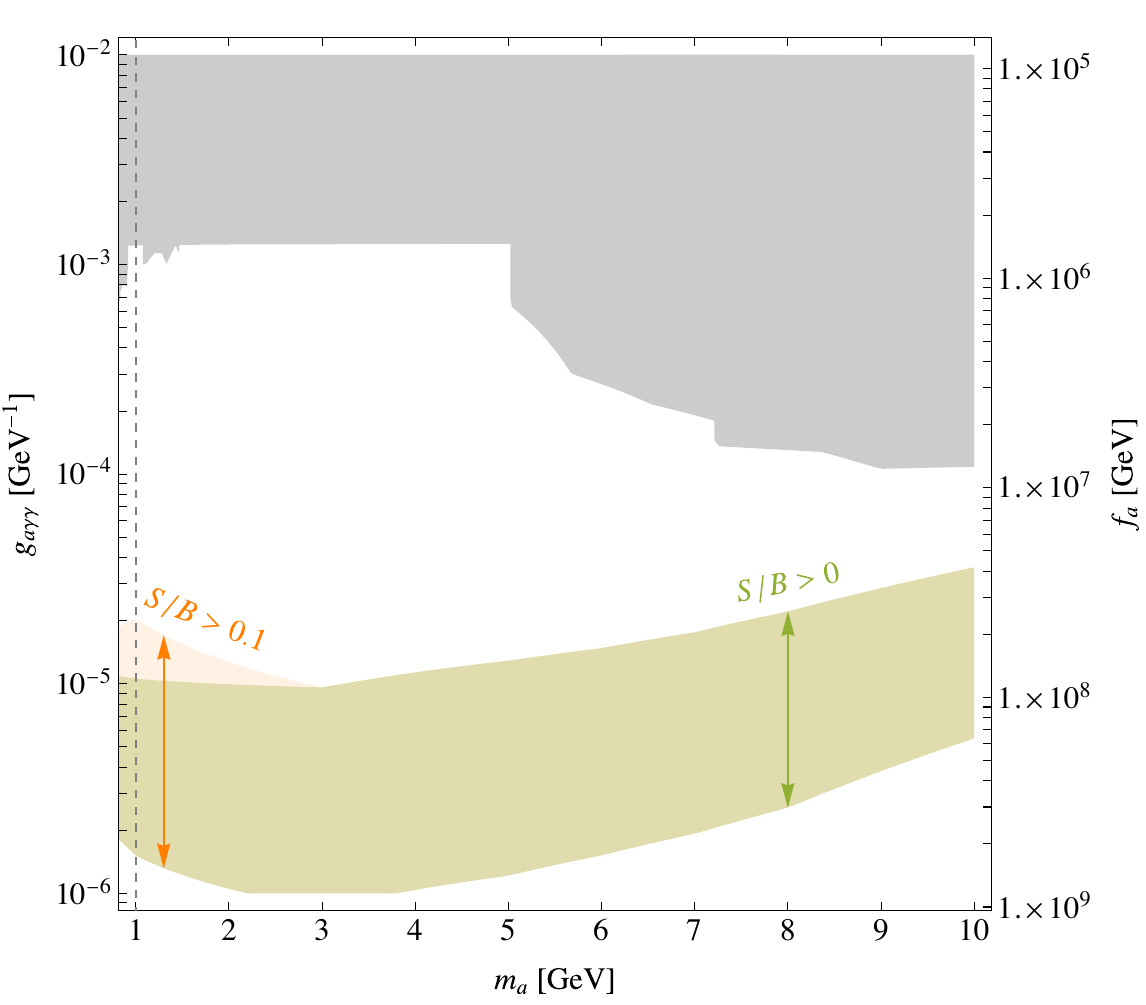}
\caption{Dependence of the expected reach on the systematic uncertainties on the background rate. The {\color{defgrey}\bf gray shaded} region shows existing constraints as in Fig.~\ref{fig:money}. \textbf{Left:} The {\color{red}\bf red} band shows the expected reach of $e^+e^-+(\gamma\gamma)_{\text{res}}$  as in \cref{fig:money}, and in {\color{magenta}\bf magenta} the same reach with the requirement $S/B>0.1$, as detailed in \cref{sec:likelihood}. The arrows of the same color show how the uncertainty of our expected reach gets larger with larger systematic uncertainties.
\textbf{Right:} The {\color{green}\bf green} band corresponds to the reach of Belle-fwd as defined in Eq.~\eqref{eq:bellefwd} and detailed in Sec.~\ref{sec:forwardfactory}; the {\color{orange}\bf orange} band is the same reach with the requirement $S/B>0.1$.}
\label{fig:sys}
\end{figure}

The shape of the likelihood distribution can be explained by noticing that the absolute values of the variables $t_{\pm}$ as defined in Eq.~\eqref{eq:definitions} are bounded from below by a quantity that depends on the detector acceptance which determines the minimal polar angle of the electron/positron that one can measure, $\theta_{e_{out}^{\pm}}^{\text{min}}$. In formulas we can write the bound for tip of the wedge with $t_{+}~\sim t_{-}$ as
\begin{equation}
\vert t_{\pm}\vert>\frac{s}{4}(1-\cos\theta_{e_{out}^{\pm}}^{\text{min}})\ ,
\end{equation}
where $\theta_{e_{out}^{\pm}}^{\text{min}}$ can be extracted from Eq,~\eqref{eq:BelleIIopen} for Belle~II and from Eq.~\eqref{eq:bellefwd} for Belle-fwd. The comparison between the two panels shows clearly how the extended forward coverage opens up a new region of phase space where the likelihood is considerably larger, hence boosting the sensitivity to the ALP.   

We now illustrate the dependence of the likelihood on our treatment of the empty bins illustrated in Appendix~\ref{sec:MC}. Calling $S$ the expected signal events  and $B$ the expected background events, we show in Fig.~\ref{fig:ll_syst_triplet} the distribution of the likelihood in the plane spanned by $S/\sqrt{B}$, which roughly controls the statistical weight, and $S/B$, which is a good proxy for the sensitivity of our result on the systematic uncertainties in the background rate. The three panel illustrate the three different treatments of the empty bins, and the coupling strength of the ALP in each panel is fixed at its corresponding expected sensitivity at Belle~II. 

In the left panel we show the likelihood assuming that the probability distribution of the empty bins is Poissonian and has its mean set at the upper bound of the background cross-section obtained from our MC exploration. The discrete nature of the likelihood makes most of the likelihood concentrated on the tail of the Poisson distribution with zero background events which is shown as a red square dot in the plot. The center panel shows the likelihood obtained under the assumption that the events in the empty bins are uniformly distributed with maximal value given by the upper bound derived in our MC. Here we see that the likelihood distribution becomes nicely continuous in the $(S/\sqrt{B},S/B)$ plane. In the right panel we show the likelihood distribution where we assume the background rate in the empty bins to be fixed at the lower bound derived from the MC. This assumption gives the most conservative expected sensitivity in Fig.~\ref{fig:money}. For both the center and right panel, we observe that a large fraction of the likelihood lives at large value of $S/B$, anticipating the low dependence of our expected reach on systematic uncertainties on the background.

The dependence of the expected reach on the systematic uncertainties is shown in Fig.~\ref{fig:sys}. To quantify this in our likelihood analysis we restrict our attention to the bins with $S/B$ larger than a given value. The lower bound on $S/B$ can be taken as a proxy for the expected systematic uncertainty. For $S/B>0.1$ we see that the uncertainty in our expected reach is sensibly increasing at Belle~II (left panel) while it is left almost unchanged at Belle-fwd (right panel). The sensitivity of the likelihood to the cut in $S/B$ depends very much on the statystical treatment of the empty bins as can bee inferred from Fig.~\ref{fig:ll_syst_triplet}. In particular, for the  we assume the events in the empty bins to be distributed as a Poissonian the likelihood is dominated by bins with $B=0$ and therefore essentially insensitive to systematic uncertainties. This is confirmed by looking at the left panel of Fig.~\ref{fig:ll_syst_triplet}. Therefore, accounting for systematic uncertainties does not shift our best sensitivity. Conversely, systematic uncertainties affect our most conservative sensitivity which is obtained by taking the upper bound on the cross-section derived in our MC as the number of expected background events. At Belle~II we see how having systematic uncertainties at 10\% level shifts the reach by a roughly a factor of five on the coupling. This is confirmed by looking at the right panel of Fig.~\ref{fig:ll_syst_triplet}.

\section{Effective photon approximation at flavor factories}\label{app:EPA}

The photon fusion process that we propose as a new channel to search for ALP production in \BelleII is traditionally linked to the equivalent vector boson approximation~\cite{Fermi:1925fq,Williams:1934ad,Budnev:1975poe,Frixione:1993yw}. In the context of LHC physics this approximation provides a simple computational tool  to estimate the total cross-section of fusion processes, see e.g. Ref.~\cite{Dawson:1984gx,Cahn:1983ip} for early applications in the context of Higgs boson production in hadron colliders.
While the computational aspect is no longer as pressing as it used to be, the equivalent boson picture can in principle be used to guide the isolation of the fusion signal from the backgrounds.

\begin{figure}[t!]
\centering
\includegraphics[width=0.6\linewidth]{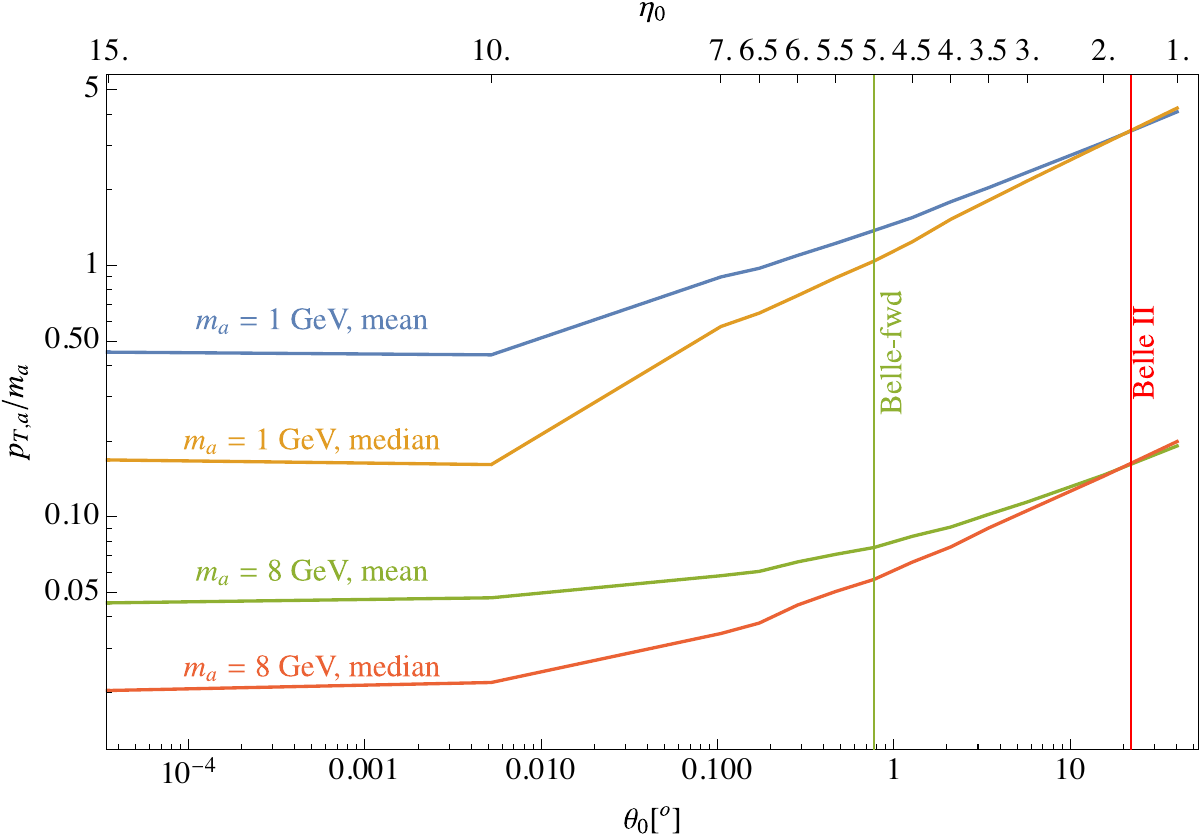}
\caption{Mean and median $p_{T,a}/m_a$ for \cref{eq:signal-intro} including phase-space up to a minimal angle $\theta_0$ (maximal pseudo-rapidity $|\eta_0|$).}
\label{fig:mean_median_pt_1_8_gev}
\end{figure}

The basic idea of the approximation is to consider as  on-shell the gauge boson field that enters in a scattering, as long as its virtuality can be neglected in comparison with the momentum exchanged in the scattering in which the boson participates. In our case the photon, in order to be approximately on-shell, must have negligible virtuality with respect to the mass of the $a$ that emerges from the photon-fusion. The physical foundation of this approximation relies on the fact that the process of photon fusion happens in a time of order $1/m_a$, which is too short to probe the virtuality of the photon, $q_i^2$ as defined in Fig.~\ref{fig:ee_eea_tchan_diagram} if $q_{i}^2\ll m_a^2$~\cite{Borel:2012by}. The virtuality of the photon can only be ``detected'' in experiments sensitive to much longer time-scales.

As in our kinematics the photon virtualities are linked to the final state electron and positron transverse momenta 
\begin{equation}
q_{1}^2\simeq p^2_{T,e_\text{out}^-},\quad q_{2}^2\simeq p^2_{T,e_\text{out}^+} \quad \text{ and } \quad p_{T,e_\text{out}^+} + p_{T,e_\text{out}^-}=p_{T,a}\ ,
\end{equation}
it is instructive to look at the typical transverse momentum of the ALP divided by its mass. In 
\cref{fig:mean_median_pt_1_8_gev} we show the median and the mean $p_{T,a}/m_a$ for $m_a=1\text{ GeV }$ $m_a=8\text{ GeV }$ including phase-space up to a minimal angle $\theta_0$ (maximal rapidity $|\eta_0|$) from the beam axis. We observe that for 1 GeV ALP mass, and for similar ALP masses compared to the center of mass energy, $p_{T,a}/m_a\sim 1$, which indicates that such light ALP is not accurately described by the equivalent photon approximation. This is especially true for $\theta_0\simeq 1\degree$ and $22\degree$, the benchmarks for the Belle-fwd hypothetical detector and the actual \BelleII detector respectively. 

\begin{figure}
\centering
\includegraphics[width=0.49\linewidth]{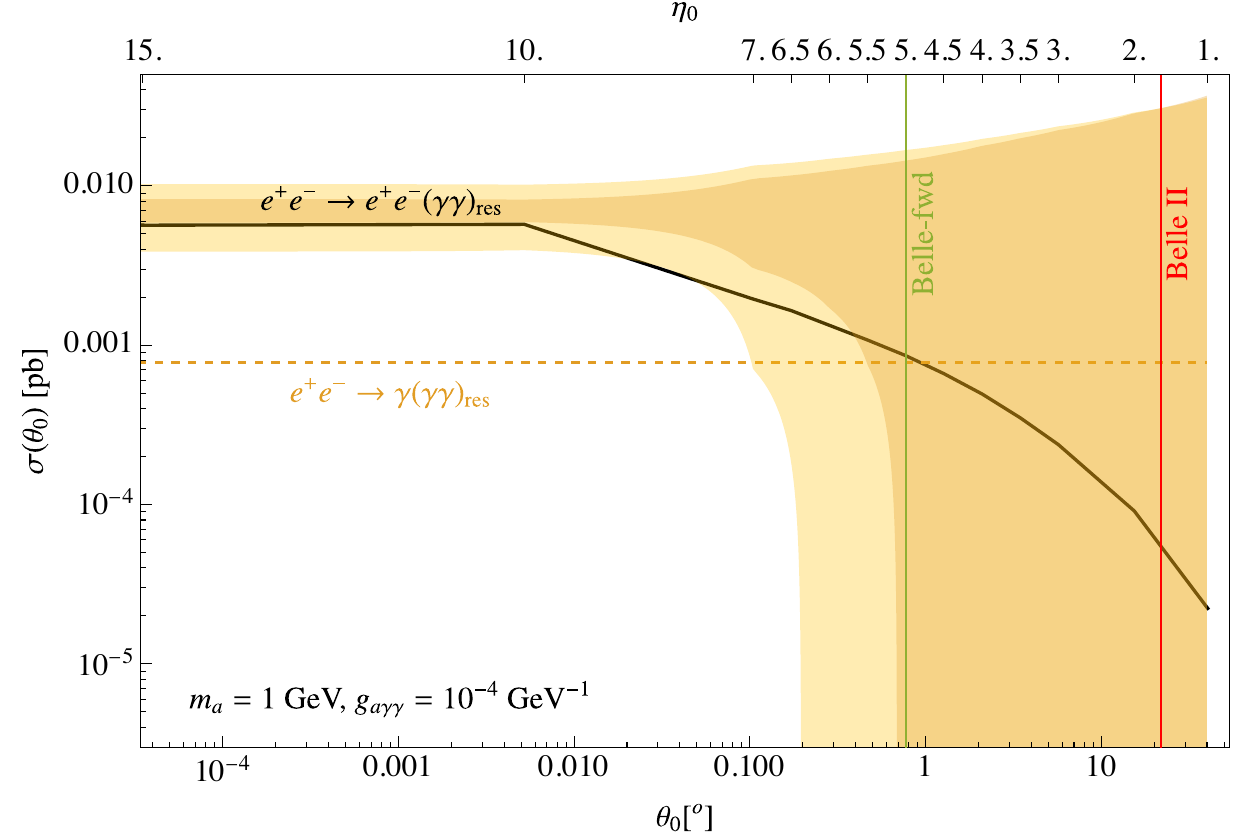}\hfill
\includegraphics[width=0.49\linewidth]{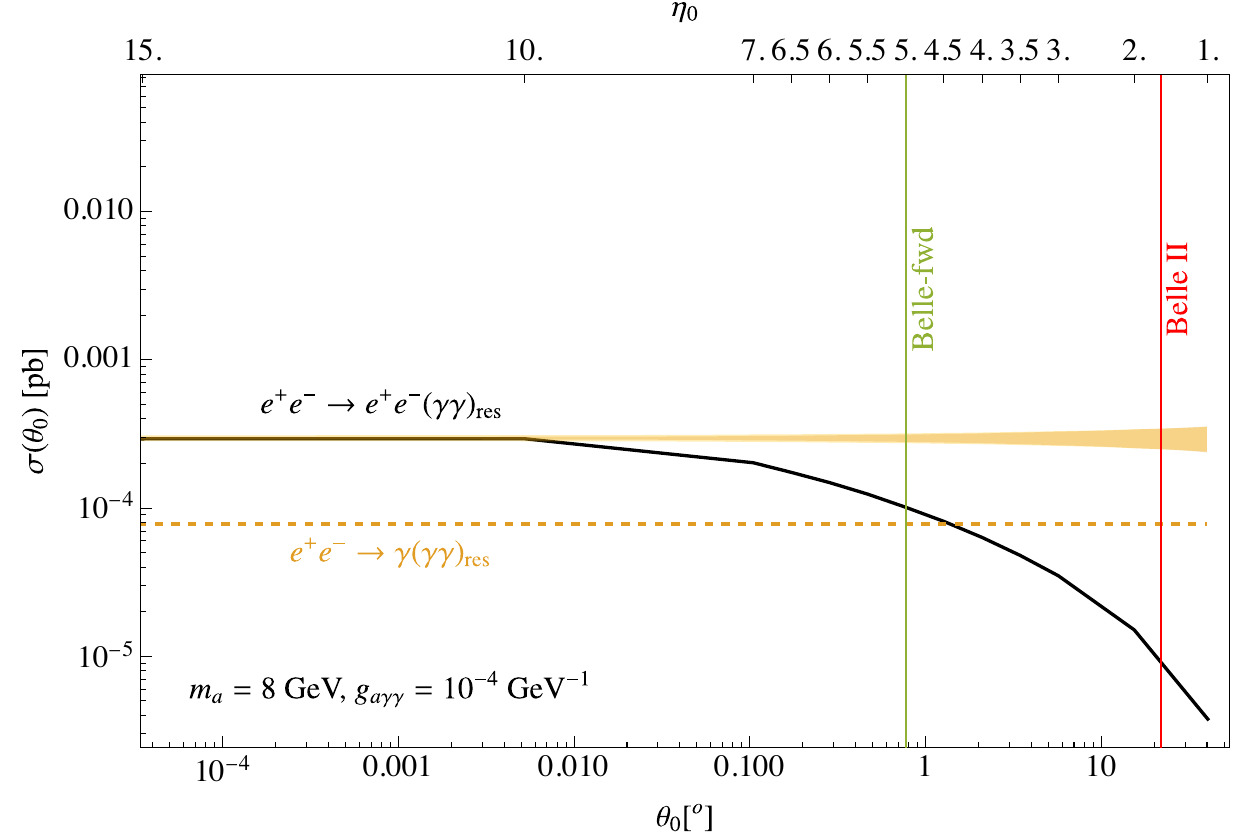}
\caption{Comparison of the EPA prediction and the exact cross-section   for the process \cref{eq:signal-intro}, in the case of $m_a=1$~GeV (left) and $m_a=8$~GeV (right). The \textbf{black} line is the cross-section for the exact computation of the process \cref{eq:signal-intro} as a function of the lepton acceptance (see \cref{eq:BelleIIopen}), expressed either as a polar angle measured in degrees $\theta_0$ or as a pseudo-rapidity $\eta_0$. The {\color{mean}\bf light yellow} and the {\color{median}\bf dark yellow} error bands are the EPA with mean and median $p_{T,a}/m_a$ relative error, as detailed in the text. The vertical {\color{red}\bf red} and {\color{green}\bf green} lines correspond to the angular acceptance at Belle~II and Belle-fwd respectively. For reference we also show the   $\gamma+(\gamma\gamma)_{\text{res}}$ cross-section (which is invariant with respect to $\theta_0$) as an {\color{orange}\bf orange dashed} line.} 
\label{fig:EPAconvergence}
\end{figure}

To corroborate this expectation, we show in the left panel of \cref{fig:EPAconvergence} the exact calculation of the fusion process \cref{eq:signal-intro} and the EPA prediction as computed at its lowest order following the definition of Ref.~\cite{Budnev:1975poe}. The EPA prediction is further  equipped with an error band in light (dark) yellow, which corresponds to an uncertainty given by the mean (median) $p_{T,a}/m_a$ shown in \cref{fig:mean_median_pt_1_8_gev}.
The agreement between the EPA and the exact computation degrades as $\theta_0\gtrsim 0.01\degree$ ($|\eta_0|\lesssim 10$), with an order of magnitude disagreement expected for Belle-fwd and even worse for Belle~II. In the right panel of \cref{fig:EPAconvergence} we show the same quantities for a heavy ALP mass of 8~GeV. In this case a much better agreement between EPA and the exact computation is observed, in full compliance with the smallness of the ratio $p_{T,a}/m_a\lesssim 0.1$ over the whole range of $\theta_0$.

From these analyses we conclude that the EPA works well in the cases in which the observed final states corresponds to small $t$-channel photon virtualities, but it is definitively possible to reach situations in which EPA does not provide a good estimate of the ALP total production cross-section. In these situations, also the kinematic picture of a signal dominated by small $p_{T,a}$ can be flawed. In fact, for ALPs mass  1 GeV in \cref{fig:tplus-tminus-splus-sminus} we have observed a strong preference for large $E_{\gamma_1} + E_{\gamma_2} \gg m_a$, in spite of the EPA expectation of an ALP produced nearly at rest. 

Due to the difficulty to build forward detector in machines capable of reaching high instantaneous luminosity, such as flavor factories, the EPA should be used with great care to compute cross-sections at these machines. Similarly, the physics intuition that follows from the validity of the EPA can be inaccurate for  experiments on these machines, due to the absence of coverage in the forward phase-space.

\section{Influence of ALP lifetime}
\label{app:lifetime}
For small enough ALP masses, the decay length becomes so large that most of the time the ALP decay happens far away from the interaction point, possibly even outside the detector. In order to account for this effect, we reweigh every event by including the probability of the ALP to decay outside the detector length. The decay probability before some distance $L$ can be written as
\begin{equation}
P_{\text{decay}}=1-e^{-\frac{L}{\gamma_a \tau_a}} = 1-e^{-\frac{L \Gamma_a m_a}{E_a}}\ ,
\end{equation}
where  $m_a$ and $E_a$ are respectively the ALP mass and energy and 
\begin{equation}
\Gamma_a= \frac{\gagg^2 m_a^3}{64\pi}
\end{equation}
its decay width. In order to make the most conservative estimate, we will choose $L$ to be the shortest distance between the interaction point and the ECL barrel which is $L=1.25 \text{ m}$ at Belle~II~\cite{Belle-II:2018jsg}. Each event will then have a weight which is a function of $\gagg$ and also depends on the ALP mass and energy. We can then include this effect in the likelihood of Eq.~\eqref{eq:likelihood}. The effects of the ALP lifetime, as expected, are visible only for $m_a\lesssim 0.5 \GeV$; for those masses there is effectively a reduction of the signal rate. As shown in \cref{fig:money}, however, for ALP masses below $1 \GeV$, invisible searches become quickly more competitive, filling the gap of the visible ones.

Interestingly, the loss of promptly decaying events is less rapid at Belle-fwd (defined in Eq.~\eqref{eq:bellefwd}) with respect to Belle~II (see Eq.~\eqref{eq:BelleIIopen}). This can be explained by computing the average boost of the particle as a function of the opening angle of the detector. For a more forward detector like Belle-fwd the average boost of the ALP will be smaller, making it easier for the ALP to decay inside the detector.

\bibliographystyle{JHEP}
\bibliography{bib}

\end{document}